\documentclass[useAMS,usenatbib]{almamemo}
\pdfoutput=1
\usepackage{amsmath}
\usepackage{microtype}
\usepackage{mathptmx}
\usepackage{graphicx}
\usepackage{booktabs}
\usepackage{tabularx}
\usepackage[colorlinks=true,citecolor=blue]{hyperref}
\usepackage[]{subfig}
\usepackage{astrojournals}

\newcommand{\wvropt}[1]{{\tt --#1}}
\newcommand{\wvrgcal}{\texttt{wvrgcal}}

\newcommand{\code}[1]{\texttt{#1}}
\newcommand{\dir}[1]{\texttt{#1}}

\volume{593} \title[WVRGCAL]{Design and Implementation of the {\tt
    wvrgcal} Program}

\author[Nikolic et al]{B. Nikolic,  S.~F. Graves, R.~C. Bolton and J.~S. Richer \\
  Astrophysics Group, Cavendish Laboratory, Cambridge CB3 0HE, UK
  \\\url{email:b.nikolic@mrao.cam.ac.uk}
  \\\url{http://www.mrao.cam.ac.uk/~bn204/}}

\pagerange{\pageref{firstpage}--\pageref{lastpage}; } \pubyear{2012}

\begin{document}
\label{firstpage}

\maketitle

\begin{abstract}
  This memo describes the software engineering and technical details
  of the design and implementation of the \wvrgcal\ program and
  associated libraries.  This program performs off-line correction of
  atmospheric phase fluctuations in ALMA observations, using the
  183\,GHz Water Vapour Radiometers (WVRs) installed on the ALMA 12\,m
  dishes.  The memo can be used as a guide for detailed study of the
  source code of the program for purposes of further development or
  maintenance.
\end{abstract}

\section{Introduction}

The `{\tt wvrgcal}' program is an application for off-line correction
of atmospheric phase fluctuations in ALMA data based on observations
of 183\,GHz Water Vapour Radiometers (WVRs) that are installed on all
ALMA 12\,m-diameter antennas. The principles of WVR based phase
correction, and of the algorithms used in `{\tt wvrgcal}' are
described in previous papers and ALMA memos, e.g.,
\cite{2001ApJ...553.1036W,ALMANikolic587,ESONikolic2008}, and
forthcoming publications. In this memo we describe the technical
details of the software engineering design and implementation of the
{\tt wvrgcal} program. This memo describes version 1.2 of \wvrgcal\,
which can be downloaded under the GNU Public License at
\url{http://www.mrao.cam.ac.uk/~bn204/soft}.

Brief usage instructions for astronomical users of \wvrgcal\ can be
found in the command line help (\texttt{wvrgcal --help}), shown in
Appendix \ref{sec:help}. The program is currently shipped with the
data reduction environment CASA (from version 3.4), and
additional usage instructions can be found in the integrated help
within CASA and in the CASA cookbook
(\url{http://casa.nrao.edu/ref_cookbook.shtml}).

\section{Design}

\begin{figure*}
  \includegraphics[width=\textwidth]{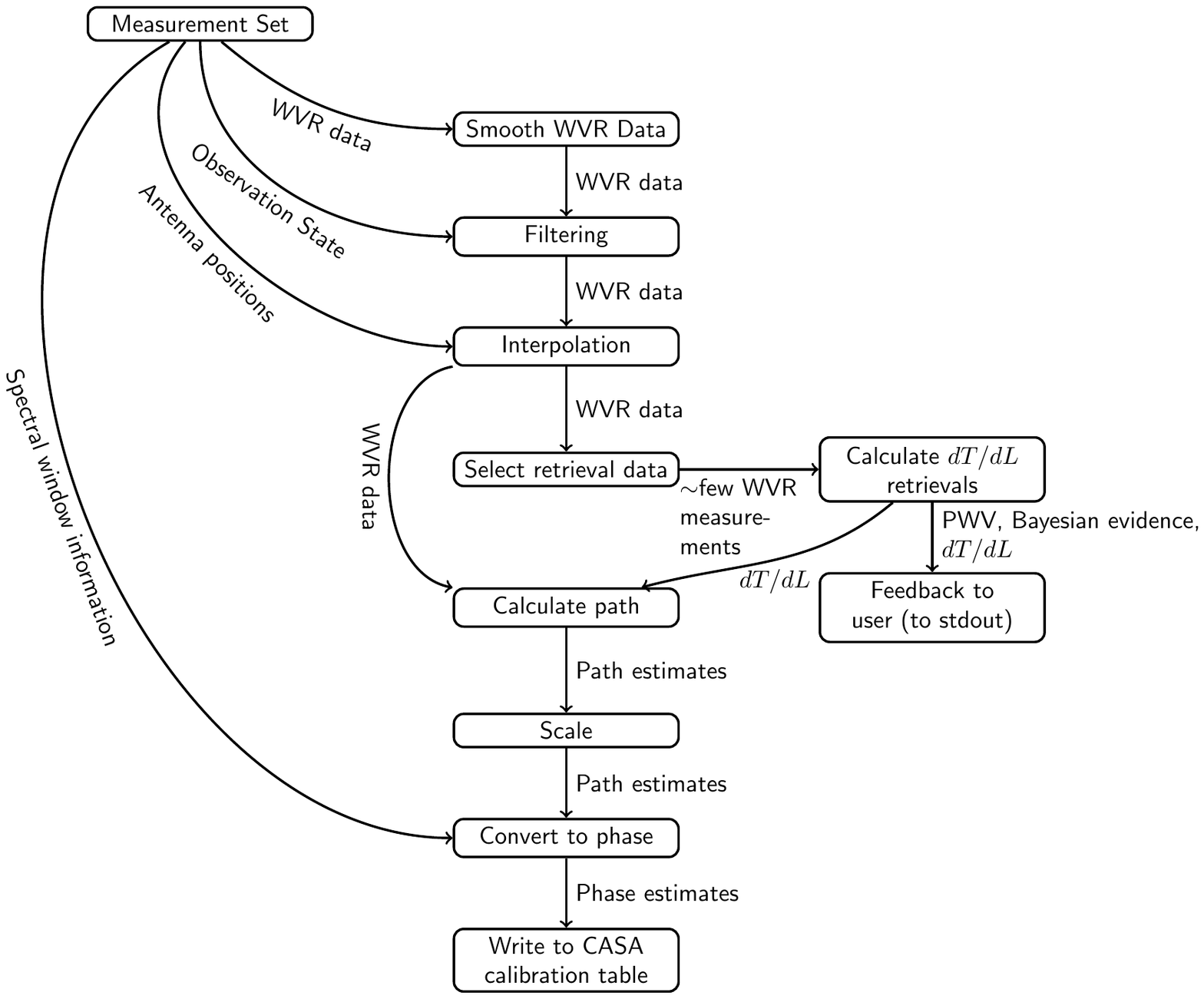}
  \caption{Flow of information within the {\tt wvrgcal} program} 
\end{figure*}

The phase correction process consists of four conceptual stages:
\begin{enumerate}
  \item Computation of the \emph{phase correction coefficients}.
  \item Use of the phase correction coefficients to transform the
    fluctuating sky temperatures into path estimates.
  \item Transformation of the path estimates to channel-specific phase corrections.
  \item Application of the phase correction to the astronomical visibility data.
\end{enumerate}
The {\tt wvrgcal} program handles the first three stages of this
process. The final stage is carried out similarly to the application
of other calibrations -- e.g. by using  CASA or another
interferometric data reduction package to apply the calibration table
produced by \texttt{wvrgcal}.

The design and control flow through the program consist of the first
three stages, with a user-interface and control layer on top of them.

\subsection{Initial processing of WVR data}
\label{sec:initprocess}
Before the main phase correction steps are done, there is an initial
pre-processing stage during which possible problems or artifacts are removed
from the WVR data and optional smoothing is applied. Currently this
processing consists of:
\begin{enumerate}
\item Filtering out measurements which may be affected by ALMA
  Calibration Arm  blocking the WVR beam. 
\item Replacing data antennas with \emph{flagged\/} WVRs or with no
  WVR at all with measurements interpolated from nearby antennas.
\item Smoothing the WVR measurements in time.
\end{enumerate}

The filtering stage is carried out first to ensure that no averaging
is done over data that may be contaminated. The filtering works on the
basis of `scan intent' of each of the measurements. Specifically, only
WVR data marked with {\tt ON\_SOURCE} and \emph{not marked\/} with
{\tt CALIBRATE\_ATMOSPHERE} intents are kept and the remaining data are
discarded. This process for example excludes any measurements when the
ALMA Calibration Device (ACD) may have been moving (which can block or
interfere with the WVR beam) or any sky-dip measurements of
atmospheric transparency.

The interpolation stage identifies the antennas that do not have a WVR
(or have been flagged using \texttt{--wvrflag}). The WVR data for
these antennas is replaced with a data-set interpolated from the three
nearest antennas with un-flagged WVR data
(see Sec.\ref{sec:interp}). 

Smoothing of the WVR data is only carried out if the \texttt{--smooth}
option is given, along with a specified number of time samples to smooth
over. If it is requested, the smoothing process is carried out before any
interpolation or filtering is performed. The smoothing process will
not smooth across boundaries defined by the  \texttt{STATE\_ID}
flag.

\subsection{Phase correction coefficient calculation}

In the current version of {\tt wvrgcal}, each set of phase correction
coefficients is computed from a single integration (i.e., a single
time point) of WVR data from a single antenna (an important
investigation for the future is examining the differences between
coefficients on different antennas or baselines). The absolute values
of the WVR data from all four channels are used to best constrain the
inference of properties of the atmosphere. These four values and the
elevation of the antennas are currently the only information used in
the inference -- the ground level meteorological instrument data and
oxygen sounder measurements are currently \emph{not} used.

Depending on the mode of operation of {\tt wvrgcal}, one or more sets
of phase correction coefficients may be computed for each
observation. The mode of operation should be chosen by evaluating the
likelihood that the optimal phase correction coefficient could have
changed appreciably during observations. For example, if the
observation being analysed consists of two sources at very different
elevations, it is likely that the optimum phase correction
coefficients will be different for the two sources, and at least two
sets of coefficients should therefore be computed.

The {\tt wvrgcal} command line  options which control the number of phase correction
coefficient sets computed (and also how they are used) are:
{\tt --nsol}, {\tt --segsource} and {\tt --tie}.

The {\tt --nsol} option allows the user to specify directly the number
of sets of phase correction coefficients that are to be
computed. These coefficients are always computed from the first
antenna in the array, and with {\tt --nsol}
they are computed at time points evenly
distributed in time over the observation.  The {\tt --nsol} option
cannot be used together with the {\tt --segsource} or {\tt --tie}
options.  Since this option distributes the computation time points
uniformly without particular regard to the sequence of the observation
being analysed, it is usually useful only for very long tracks on a
single object.

Options {\tt --segsource} and {\tt --tie}, on the other hand, allow
selection of when to recompute the phase correction coefficients based
on the boundaries defined by the telescope moving to a new
source. These two options are normally used in combination;
furthermore, the option {\tt --tie} cannot be used if {\tt
  --segsource} is not also specified.  The effect of the {\tt
  --segsource} option is to specify that the phase correction
coefficients should be re-computed each time the telescope moves to
the next source. Here, source is used in the technical sense to mean the value in the
``SOURCE'' column in the data table. 

The explanation of option {\tt --tie} needs to be divided into two
stages, to prevent possible confusion in case of eventual further
versions of  {\tt wvrgcal}:
\begin{enumerate}
\item Conceptually, the {\tt --tie} option specifies to the {\tt
    wvrgcal} program that best effort should be made to mutually
  accurately calibrate the phase of sources specified together after
  this command line option. In other words, the phase of these sources
  is ``tied'' together as best possible. 
\item In the current version of {\tt wvrgcal}, this option in
  combination with {\tt --segsource} is {\bf interpreted} to mean that
  changes between sources in a ``tie'' group are \emph{not\/} a
  trigger for re-calculation of the phase-correction coefficients. In
  other words, phase correction coefficients are re-calculated only 
  when the telescope observations moves to a source which is
  \emph{not\/} in the same tie group as the previous source.
\end{enumerate}
Typical usage of this option is that science targets are ``tied''
together with their phase calibration sources, but not,for
example, with the amplitude calibration sources.

Regardless of which mode is used, the time points selected for
computation are adjusted in the following way to maximise the
appropriateness of the estimated coefficients:
\begin{itemize}
\item A time point near the middle of segment of time is selected to
  make the phase correction set as representative as possible. For
  example, if only one set of coefficients is calculated for an
  observation, then that set is computed from a time sample in the middle of
  the observation (and not at the beginning).
\item The scan intent of the data point is examined to ensure the
  telescope was not doing a calibration step at the time the datum was
  recorded.
\end{itemize}

Once a WVR datum is selected for computation of the phase correction
coefficients, it is passed (together with the telescope elevation) to a
routine which does the computation using Bayesian model fitting. The
technique used in the current version of {\tt wvrgcal} is as described
by \cite{ALMANikolic587}, with an improvement in that Nested Sampling
\citep{Skilling2006} is used instead of traditional Markov-Chain Monte
Carlo. The outputs of the Bayesian analysis are:
\begin{itemize}
  \item Probability distributions of the atmospheric water vapour column,
    pressure and temperature.
  \item Probability distributions of the phase correction coefficients for
    each channel of the WVR.
  \item A Bayesian evidence value, which measures how well the model
    describes the data.
\end{itemize}

Currently the only quantities used for actual calibration are the mean
values of the distributions of phase correction coefficients. The water
vapour column, the estimated error and the Bayesian evidence are
not directly used but are printed as feedback to the user. These
quantities, as well as the estimated errors on the retrieved phase
correction coefficients, could be used in the future to improve the
phase correction performance -- for example, by not using WVR channels
for which the phase correction coefficients are poorly estimated.

Once the phase correction coefficients are computed, they are used,
together with the WVR data, to estimate the phase fluctuation for each
antenna. In the standard mode of operation the path is computed 
as  a weighted linear combination of the WVR brightness temperatures in the four
channels:
\begin{equation}
  \Delta L(t) = \sum_{k=1}^4 w_k T_{{\rm B},k}(t).
\end{equation}
There is also an experimental (but not commissioned) mode of {\tt wvrgcal}
where the calculation is second order in path with respect to observed
sky brightness.

When a single set of phase correction coefficients is used for the
whole observation, the weights are given by:
\begin{equation}
  \frac{1}{w_k}= \left( \delta T_{{\rm B},k}   \frac{{\rm d} L}{{\rm d} T_{{\rm
          B},k}}\right)^2 
  \sum_{i=1}^4 \left(\frac{1}{\delta T_{{\rm B},i}   \frac{{\rm d} L}{{\rm d} T_{{\rm  B},i}}}\right)^2.
\end{equation}
Where:
\begin{description}
\item[$\mathrm{d} L$:] Final path estimate 
\item[$w_k$:] The weight of the $k$-th channel
\item[$T_{{\rm B},k}$:] Observed brightness temperature of the $k$-th
  channel
\item[$\delta T_{{\rm B},k}$:] Expected thermal-like noise in the
  $k$-th channel of WVRs
\item[$\frac{{\rm d} L}{{\rm d} T_{{\rm
          B},k}}$:] Phase correction coefficient of the  $k$-th channel
\end{description}
When multiple sets of phase correction coefficients are used, the
values of $\frac{{\rm d} L}{{\rm d} T_{{\rm B},k}}$ to be used in
equation 2
 are always linearly interpolated in time between the
two nearest sets of coefficients. Segments of time over which
\emph{fixed} coefficients are desired can be specified by entering the
same coefficients at the beginning and end of the time segment -- in
this way the linear interpolation has no effect. This is the technique
used when multiple coefficients are computed using the {\tt
  --segsource} flag.

At this stage, if any sources are flagged with \wvropt{sourceflag},
then the path estimates corresponding to the times when these sources
were observed are set to zero. In this way no phase corrections for
these sources is applied. This is useful when the
observations of these particular sources are known to be corrupted,
e.g., due to shadowing (one antenna
blocking  part of the field of view of another antenna).

The next stage is to compute and provide some feedback to the
user. This feedback consists of the RMS of the path fluctuation for each
antenna and also  the path `discrepancy', which is described
in \cite{ALMANikolicDiscrepancy}. In order to make these statistics
easier to interpret, the \wvropt{statsource} option can be used to
compute the statistic for only a subset of the entire
observation. When this option is selected, time intervals
corresponding to the specified sources are computed and only the path
estimates falling into these time intervals are used to compute the
statistics.

After computing the user feedback statistics, the estimated paths can
be further scaled by using the \wvropt{scale} parameter. This
parameter \emph{can} be used to fine-tune the magnitude of the correction, but
currently it is not recommended as the best way to use this has not
been thoroughly studied.

As the final part of the processing, the details of the spectral setup
of the astronomical receivers is loaded and the computed path
estimates are converted to a phase estimate. The conversion is carried
out separately for each astronomical spectral window and channel, to
take into account the changing wavelength. At this stage a dispersion
correction can also be made (option \wvropt{disperse}), but this mode
is also not commissioned and therefore not recommended for use.

\section{Implementation}

\subsection{Build system}

The compilation of the LibAIR package and the {\tt wvrgcal} program is
implemented using the standard GNU autoconf/automake build system. The
benefits of this build system are:
\begin{itemize}
  \item Very high level of standardisation -- the majority of GNU/Linux
    libraries and applications are built using this system.
  \item Detection of features of the host system and checks for
    presence of prerequisite libraries.
  \item Automatic dependency tracking for C/C++ programs.
  \item Facilities for running unit-tests and creation of distribution
    tar-balls.
\end{itemize}

The {\tt configure} script is generated from the {\tt configure.ac}
specification. The configure script takes a number of options that
modify the way the LibAIR package is built. Some of the important
options are:
\begin{description}
  \item[\wvropt{disable-pybind}] Do not try to build the Python
    binding. This option is recommended for all builds which are not
    intended for in-depth development of the library.
  \item[\wvropt{with-casa}] Location of the CASA package, which is
    required for building {\tt wvrgcal}. The directory should point to
    the top of the CASA directory tree.
  \item[\wvropt{disable-buildtests}] Do not build the unit-tests --
    saves some time in compilation and removes the dependency on the
    Boost Unit Test library.
\end{description}
There are detailed worked examples on how to build the LibAIR library
and all pre-requisites available at
\url{http://www.mrao.cam.ac.uk/~bn204/alma/sweng/libairbuild.html},  and
pages referred from there.

\subsection{External Libraries} 

\begin{figure*}
  \includegraphics[width=1\linewidth,trim=120 450 0
  130]{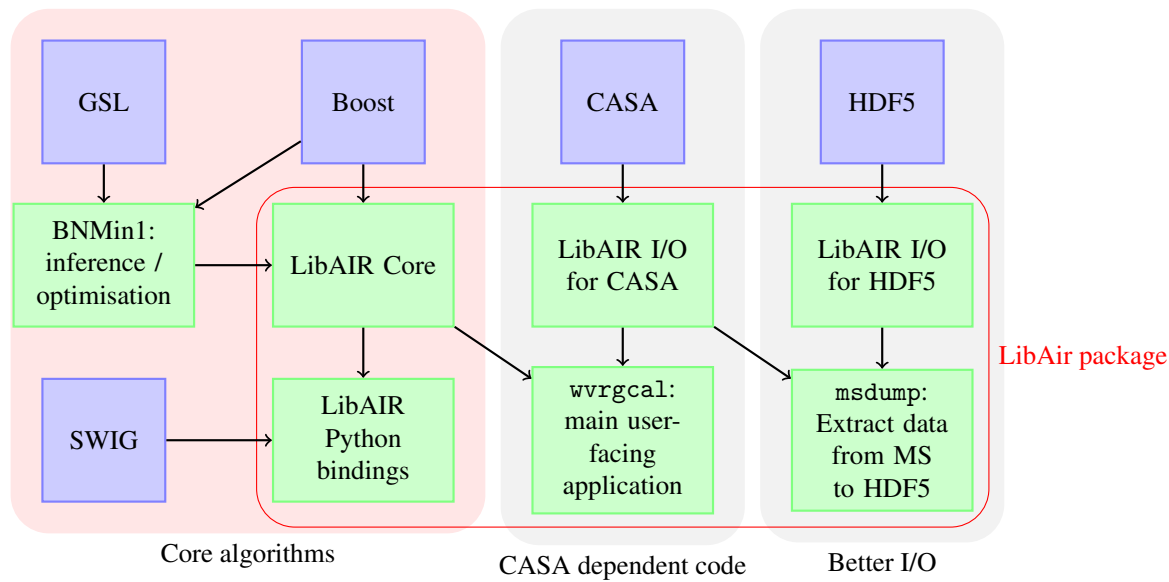}
  \caption{Illustration of the software architecture of the {\tt
      wvrgcal} program and supporting libraries.}
  \label{fig:libraryarch}
\end{figure*}

The LibAIR package and {\tt wvrgcal} program make use of a number of
external libraries. The architecture, which is illustrated in
Figure~\ref{fig:libraryarch}, has been designed to minimise
cross-dependencies between non-standard external libraries.

Here is a short description of each of the external libraries:
\begin{description}
  \item[{\bf Boost Libraries:}] The Boost libraries are used for a
    variety of algorithms, containers and utilities throughout the
    package. They are a prerequisite for the entire LibAIR package and
    LibAIR can not be compiled without them. The Boost libraries can
    be downloaded from \url{http://www.boost.org/}.
  \item[{\bf BNMin1:}] This is a minimisation/inference library. It is
    used to compute the Bayesian fit of the model to the observed
    atmospheric brightness in WVR channels. The library is available
    for download at
    \url{http://www.mrao.cam.ac.uk/~bn204/oof/software.html}.
  \item[{\bf GNU Scientific Library (GSL):}] is a pre-requisite of
    BNMin1 package and is not used directly in LibAIR/wvrgcal. It is
    available for download from \url{http://www.gnu.org/software/gsl/}.
  \item[{\bf HDF5:}] This is an optional input/output library for
    binary data. It is optionally used by {\tt wvrgcal}, by the {\tt
      msdump} program and for some testing/development purposes. The
    library is available for download from
    \url{http://www.hdfgroup.org/downloads/}.
  \item[{\bf CASA:}] {\tt wvrgcal} uses CASA
    (\url{http://casa.nrao.edu/}) libraries to read data from the
    input measurement set and write the gain calibration tables. Note
    that full CASA is required, not just casa-core
  \item[{\bf SWIG}] The SWIG program and libraries are used to
    generate the Python-language bindings to the LibAIR. These
    bindings are in turn used only for development of the library and
    therefore SWIG is not required for compiling {\tt wvrgcal} or
    other end-user tools. The SWIG program is available for download
    from \url{http://www.swig.org/}.
  \end{description}

\subsection{Directory Structure}

\begin{figure}
  \includegraphics[width=1\columnwidth]{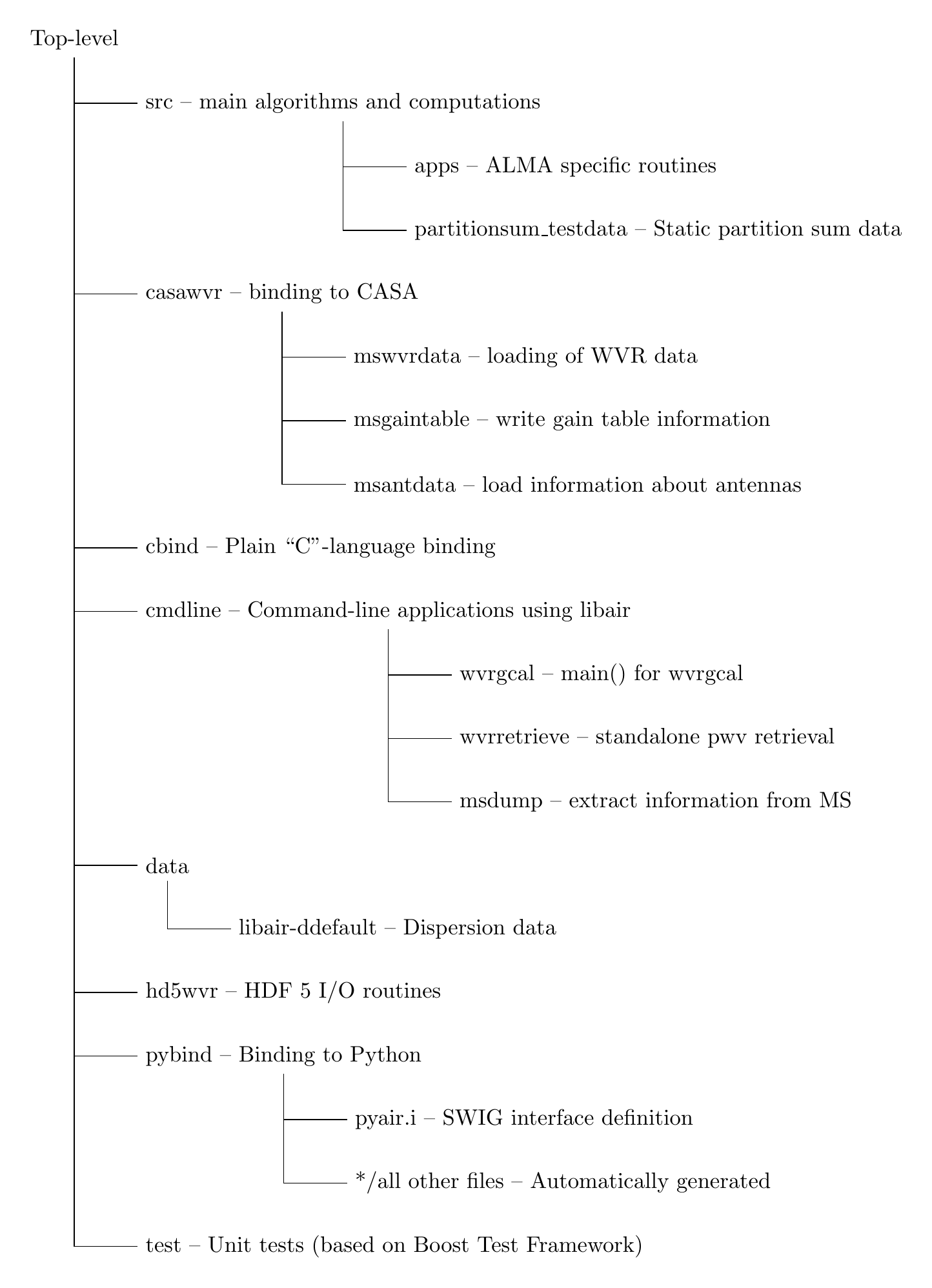}
  \caption{Directory structure of the LibAIR source code}
  \label{fig:directorystruct}
\end{figure}

The directory structure of the distributed source code is shown in
Figure~\ref{fig:directorystruct}. The code is divided up into these
directories depending on its intended functionality. The following
sections describe the various aspects of LibAIR functionality, additionally
mentioning where these can  be found within the source code tree.

\subsection{Command line parameter parsing}

The parsing of command line parameters to the {\tt wvrgcal} program is
implemented using the Boost Program Options library
(\url{http://www.boost.org/doc/libs/release/libs/program_options/}). This
is a widely used standard library for parsing command line options and
is already used in some programs packaged with the
CASA environment (e.g., the {\tt asdm2MS} program). 

The listing of available command line options and a short description
of each for \wvrgcal\ are shown in Appendix
\ref{sec:help}. These can be seen by executing \texttt{wvrgcal
  --help}. 

\subsection{Loading data from Measurement Sets}

The {\tt wvrgcal} program uses only the WVR data from the telescopes
and does not require the astronomical auto- or cross-correlation
data. As the volume of WVR data in all observations is very small it
is feasible and convenient to load all of it into memory data
structures for further processing. The memory data structure that is
used is for storing this information is {\tt InterpArrayData} which is
declared in the file {\tt src/apps/arraydata.hpp}. This data consists
of simple linear arrays into which the relevant data from the MS are
loaded. The arrays are `row-synchronous', e.g., $i$th element of the
time array is the timestamp of the $i$th element of the WVR data
array.

The loading itself, like all other code which needs to make use of
casa-core or CASA libraries, is in the {\tt casawvr} directory, 
primarily in {\tt mswvrdata.hpp}.

\subsection{Smoothing}
If the  command line option \texttt{--smooth} is called, then this
smoothing of the WVR data is done
before any other processing of the input data. 
 If requested, this option replaces each WVR data point with a mean
calculated within a smoothing window of the requested time width, and centred on that
data point. If the number of samples to smooth over is an odd number,
then the smoothing window is the requested width. If the number of
samples ($n$) is an even number, then a smoothing window of $n+1$ is
used, and the first and last samples are only given 50\% weight in the
averaging. The smoothing process leaves unchanged WVR samples which are
overly close to the start and end of the time range, such that
there would not be sufficient data samples to fill the entire smoothing window.
This process is performed by the function \texttt{smoothWVR}, defined within
\texttt{src/apps/arraydata.cpp}. 

\subsection{Data filtering}
The data filtering discussed in Sec.\,\ref{sec:initprocess} (i.e.\ only
 data where the `scan intent' is marked as \texttt{ON\_SOURCE}
but not marked as \texttt{CALIBRATE\_ATMOSPHERE} are passed
through). The identification of the correct data is done in the
\texttt{skyStateIDs} function within
\texttt{casawvr/msutils.cpp}. This information is passed by \wvrgcal{}
to the filtering function \texttt{filterState}, defined in
\texttt{src/apps/arraydata.cpp}, which filters out unwanted data from
further analysis.

A similar process is used to filter out sources flagged by the user. A
function \texttt{filterInp} is defined within
\texttt{cmdline/wvrgcal.cpp}, and is called if the
\texttt{--sourceflag} option is used; only samples whose source is \emph{not} flagged are passed back out.  This
occurs just before the coefficient calculation is performed.

\subsection{Interpolation}
\label{sec:interp}
The interpolation of WVR data for antennas with no WVRs or with bad
data is implemented in three stages:
\begin{enumerate}
\item Enumeration of antennas that need to be interpolated. This is
  done by parsing the \wvropt{wvrflag} command line option to get
  user-flagged antennas and calling the {\tt NoWVRAnts} function
  (defined in {\tt cmdline/wvrgcal.cpp}) to
  get automatically flagged antennas. {\tt NoWVRAnts} currently
  automatically flags all antennas with names that begin with letters
 `CM' (i.e., the compact array 7\,m-diameter antennas, which do not
 have WVRs).
\item Computation of nearest three antennas to each flagged antenna
  (excluding other flagged antennas) and the weight to be used
  (proportional to inverse distance) for each of the antennas
  (function {\tt linNearestAnt} in file {\tt src/apps/antennautils.cpp})
\item Replacing the data for bad antenna by the linear combination of
  data from the computed nearest antennas (function {\tt
    interpBadAntW} in file {\tt src/apps/arraydata.cpp}).
\end{enumerate}

\subsection{Atmospheric model}

One of the key parts of the LibAIR package is the forward model for
sky brightness around the $183\,$GHz water vapour line. This model is
subsequently used for both the inference of the atmospheric properties
and computation of the phase correction coefficients. The model is
defined in files within the {\tt src/} directory. 

The physics of the model is described in \cite{ALMANikolic587}. The
implementation of the model is contained in files in the {\tt src}
sub-directory of the LibAIR distribution. The C++ code implementing
the model is architectured as a polymorphic class hierarchy with the
key functions being:
\begin{itemize}
  \item {\tt eval} which computes the sky brightness in one channel
    or in all channels simultaneously
  \item {\tt dTdc} which computes the rate of change of sky brightness
    w.r.t. water vapour column
\end{itemize}
These are declared in the base class {\tt WVRAtmoQuants} in file {\tt
  model\_iface.hpp}. 

The actual model used is made up by combining two classes: starting
with a single layer water vapour model (class {\tt ISingleLayerWater}
declared in file {\tt singlelayerwater.hpp})
and then adding model for the effect of zenith angle and imperfect
coupling between WVR and sky (class {\tt CouplingModel} in {\tt model\_iface.hpp}).  Each
component of the model declares its own parameters that are changeable
at run time through the virtual function {\tt AddParams}.
  
\subsection{Implementation of retrieval of phase correction
  coefficients}

After filtering and interpolation of the WVR data, {\tt wvrgcal} next
selects the individual time samples from which the phase correction
coefficients are to be calculated. This selection is made by functions
in the \dir{src/apps/almaabs.cpp} file. If the \wvropt{segsource}
option was supplied then the selection of time samples is done in
function \code{FieldMidPointI}, which first calculates the range of
times during which each set of separate sources
(separated by function \code{fieldSegmentsTied} in \dir{src/apps/segmentation.cpp}) was
observed and then finds the mid point in time for each of these. 
Finally, it selects the next time sample after this midpoint which does not have a
`bad' (i.e., obscured by the ACD arm or not {\tt ON\_SOURCE}) state.

If the \wvropt{segsource} was \emph{not\/} supplied, then a fixed
number of coefficients, as selected by the \wvropt{nsol} option
(default is 1), will be computed. The selection of data for this is
done by function \code{MultipleUniformI}. This first finds equally
spaced points over the full time range of the observation, and then
selects the first sample after each of these time points which has a
correct state (i.e. on source, and not performing atmospheric
calibration).

Both these functions for selecting data to compute phase correction
coefficients will always take the data from the \emph{first\/} antenna
only (separate computation of coefficients for each antenna may be a
useful eventual enhancement to the program).  The selected data
samples are stored within an \code{ALMAAbsInpL} structure, which is a
list of \code{ALMAAbsInput} data objects (these are defined within
\dir{src/apps/alma\_datastruct.h}), one for each set of coefficients
that will be calculated. These structures contain all the information
used for phase correction coefficient calculation -- the antenna
number, WVR readings, elevation, time, state and source of the
samples.

The next step is the calculation of phase correction coefficients,
which is triggered by the call to the function \texttt{doALMAAbsRet}
(defined in file \dir{src/apps/almaabs.cpp}).  The inputs to
this function are the selected WVR data, i.e., the \code{ALMAAbsInpL}
structure described above. The function iterates through the \code{ALMAAbsInpL}, creating
an object of class \code{ALMAAbsRet} from each datum, which itself creates an
object of class \code{iALMAAbsRet} (defined within
\dir{src/apps/almaabs\_i.cpp}) and runs \code{iALMAAbsRet::sample}. This function sets-up and runs the nested sampler (using
the class \code{Minim::NestedS} from the BNMin1 library).  The nested
sampler is specified to run for a maximum of 10000 samples -- if the
sampler is still making progress after this many samples a warning
message is printed so that the user is aware that the sampler may not
have fully converged.

The results of the computation (i.e., the evidence, the precipitable
water vapour and estimated error, and the $\mathrm{d}T\mathrm{d}L$
coefficients for each WVR time datum) are printed to screen for each
coefficient set.

The coefficients are next converted into units of Kelvin/meter, and
are used to compute the path correction for each antenna. These
corrections are stored in an object of class \code{ArrayGains}
(declared in \dir{src/almaabs/arraygains.hpp}). This class, as well as
calculating the path lengths/phase fluctuations required to create the
gains within the output calibration table, also has various filtering,
scaling and statistical functions.

If any sources were flagged by supplying the \wvropt{sourceflag}
option, then the path correction estimates for these sources are set to
zero in the function \code{ArrayGain::blankSources}. Subsequently,
various statistics on the path lengths and the expected performance
are calculated and printed to screen for the information of the
user (see sec. \ref{sec:stats}, below).

If the \wvropt{scale} option was supplied, then the path correction
estimates are now uniformly scaled by the requested fraction, via
\code{ArrayGain::scale} function.

Finally, information about the spectral setup of the observation is
loaded using the class \code{MSSpec} (declared in
\dir{casawvr/msspec.hpp}). The function \code{loadspec} defined in
this file finds the frequency and channel information
for each spectral window, using the casacore libraries to access these
details.  If the user has requested that the sign of the correction
needs to be reversed for all the data (\wvropt{reverse}) or an
individual spectral window (\wvropt{reversespw}) this is now changed
within the output gains (using function \code{reversedSPWS} defined
within \dir{cmdline/wvrgcal.cpp}.

\subsection{Computation of user feedback}
\label{sec:stats}

The \wvrgcal\ program outputs feedback to the user based on a
statistical analysis of the WVR data recorded in the input Measurement
Set.  This feedback is  mainly derived from the path length
corrections computed previously for each antenna.

The main part of the feedback is presented as a table with one row for
each antenna in the dataset. This output consists of the number and
the name of the antenna, if it contains WVR data, if it has been
flagged with \wvropt{wvrflag}, the RMS of the path lengths for that
antenna (in micron), and the discrepancy (Disc, also in units of
micron). The RMS column shows the root-mean-square fluctuations of the
computed path corrections, calculated in function
\code{ArrayGains::pathRMSAnt}. Before computation of the RMS path
fluctuation, the effect of the airmass is
removed by multiplying the path corrections by $\sin\theta$ where
$\theta$ is the elevation of the antennas.

The discrepancy column is computed in function \code{computePathDisc}
in file \dir{cmdline/wvrgcal.cpp}, which calls
\code{ArrayGains::pathDiscAnt} . It is implemented by recalculating
 the phase coefficients so that firstly only the second
channel is used, and then only the outermost channel is used. These
coefficients are used to create two sets of estimates of the path corrections for
the entire observation, which are then differenced. The RMS of the difference is
computed.

The sections of data used for computing these statistics can be
controlled by using the command line option \wvropt{statsource}. By
default the statistics are calculated using the data for the entire
observation. However, the statistics can be computed only on one
specific source by using the \wvropt{statsource}. For example, passing
\texttt{--statsource 1939-154} restricts the data used by the
statistics computation to only those recorded while the antennas were
on source 1939-154. The function \code{statTimeMask} in file
\dir{cmdline/wvrgcal.cpp} calculates the time ranges corresponding to
the chosen source for the statistics calculation.

The \wvrgcal\ program also outputs expected performance of the
calibration, which is computed in function \code{printExpectedPerf} in
file \dir{cmdline/wvrgcal.cpp}.  The following values are computed:
\begin{itemize}
\item The `thermal error' -- the expected RMS fluctuation in
  path corrections due to the intrinsic short-term noise in the mixers
  and amplifiers in the WVRs. This is computed in function
  \code{thermal\_error} (file \dir{src/apps/arraygains.cpp}) and is
  based on the nominal noise characteristics of the WVR channels,
  i.e., 0.1, 0.08, 0.08 and 0.09\,K RMS channels 1 to 4 respectively. 
\item The largest estimated path fluctuation on a single baseline,
  computed in function \code{ArrayGains::greatestRMSBl} by
  iterating through all baselines in the dataset, and for each of
  these computing the RMS
  of the difference between the phase corrections of the two antennas forming
  that baseline.
\item The estimated RMS path error due to the errors in the
  estimation of the phase correction coefficients. This is
  computed as the greatest estimated path fluctuation on a baseline (above),
  multiplied by the fractional coefficient error. 
\end{itemize}

\subsection{Dispersion correction}

The {\tt wvrgcal} program has an option to implement dispersion
correction (\wvropt{disperse}). This option has not yet been
commissioned and therefore is \emph{not\/} recommended for use on
science data. \cite{ALMACurtis590} showed that the dispersion
correction is a very weak function of atmospheric
conditions. Therefore, in the current version of wvrgcal, dispersion
correction is implemented by interpolating a static table of
correction as a function of frequency. This static table is stored in
the {\tt data/libair-ddefault.csv} file and is computed using the
(A)ATM\footnote{Available from:
  \url{www.mrao.cam.ac.uk/~bn204/alma/atmomodel.html}} program. If
this correction is requested, it is carried out within the function
that writes out the final calibration table (\code{writeNewGainTbl})

\subsection{Output of Calibration Table}

The output of the calibration tables is implemented in the file {\tt
  casawvr/msgaintable.cpp} in function {\tt writeNewGainTbl} (the name
this function reflects the changes due to the update in the format of
gain tables in version 3.4 of CASA). The implementation uses standard
CASA C++ routines for creating and filling out calibration tables. The
created calibration table has the phase correction information stored
in the standard \texttt{CPARAM} column as a complex number (the
amplitude of the correction is always unity as \wvrgcal\ does not
currently implement amplitude correction). If dispersion correction
has been requested by supplying the \wvropt{disperse} parameter this
function will also call the dispersion correction function before
writing the phase corrections.

When writing the calibration table the history sub-table is also
populated with the exact command line invocation and version of {\tt
  wvrgcal} used to generate it.

\section*{Acknowledgements}

This work was supported by the European Commission's Sixth Framework
Programme as part of the wider `Enhancement of Early ALMA Science'
project.

\bibliographystyle{mn2eurl} 
\bibliography{../alma.bib}

\appendix
\onecolumn
\section{\wvrgcal\ inline help and command line options}
\label{sec:help}
\begin{verbatim}
WVRGCAL  -- Version 1.2

Developed by Bojan Nikolic at the University of Cambridge as part of EU FP6 ALMA Enhancement
GPLv2 License -- you have a right to the source code (see http://www.mrao.cam.ac.uk/~bn204/alma)

Write out a gain table based on WVR data

GPL license -- you have the right to the source code. See COPYING

Allowed options:
  --ms arg              Input measurement set
  --output arg          Name of the output file
  --toffset arg (=0)    Time offset (in seconds) between interferometric and
                        WVR data
  --nsol arg (=1)       Number of solutions for phase correction coefficients
                        to make during this observation
  --help                Print information about program usage
  --segfield            Do a new coefficient calculation for each field (this
                        option is disabled, see segsource)
  --segsource           Do a new coefficient calculation for each source
  --reverse             Reverse the sign of correction in all SPW (e.g. due to
                        AIV-1740)
  --reversespw arg      Reverse the sign correction for this spw
  --disperse            Apply correction for dispersion
  --cont                UNTESTED! Estimate the continuum (e.g., due to clouds)
  --wvrflag arg         Flag this WVR (labelled with either antenna number or
                        antenna name) as bad, and replace its data with
                        interpolated values
  --sourceflag arg      Flag the WVR data for this source and do not produce
                        any phase corrections on it
  --statfield arg       Compute the statistics (Phase RMS, Disc) on this field
                        only
  --statsource arg      Compute the statistics (Phase RMS, Disc) on this source
                        only
  --tie arg             Prioritise tieing the phase of these sources as well as
                        possible
  --smooth arg          Smooth WVR data by this many samples before applying
                        the correction
  --scale arg (=1)      Scale the entire phase correction by this factor
\end{verbatim}

\label{lastpage}
\end{document}